\documentstyle[12pt,aasms4]{article}
\begin{document}
	     
\title{Probability distribution of density fluctuations in the non-linear
regime}

\begin{center}
{\bf \LARGE Probability distribution of density fluctuations in the non-linear
regime}

\author{J. Betancort-Rijo}
%
%

\affil{Instituto de Astrof\'{\i}sica de Canarias, E-38200 La Laguna,
Tenerife, Spain.}
\author{M. L\'opez-Corredoira}

%

\affil{Instituto de Astrof\'{\i}sica de Canarias, E-38200 La Laguna,
Tenerife, Spain.\\
Astronomisches Institut der Universit\"at Basel. Venusstrasse 7.
CH-4102 Binningen, Switzerland (actual address).}

\end{center}

\begin{abstract}

		 We present a general procedure for obtaining the present density
		 fluctuation probability distribution given the statistics of
		 the initial conditions. The main difficulties faced
		 with regard to this problem are those related to the non-linear
		 evolution of the density fluctuations and those posed by the
		 fact that the fields we are interested in are
		  the result of filtering an underlying field  with structure
		  down to scales much smaller than that of filtering. 
		  The solution to 
		 the latter problem is discussed here in detail and the solution
		 to the former is taken from a previous work.
		 
		 We have checked the procedure for values of the rms density
		 fluctuation as large as 3/2 and several power spectra and
		 found that it leads to results in excellent agreement with those
		 obtained in numerical simulations. We also recover all available
		 exact results from perturbation theory.
\end{abstract}		 
		 

\section{Introduction}

The current model for the formation of large-scale structure is based on the
gravitational instability of small initial fluctuations in the energy density. 
These are assumed to have originated from quantum fluctuations during the
inflationary epoch. The fluctuations form a random field, which is fully
specified by the $n$-point joint probability density functions (hereafter PDFs). The one-point
PDF is a useful tool for testing the statistical character of the initial
conditions (i.e. whether they are Gaussian or others), since its evolution does not depend
on the nature of the dark matter (Trimble 1987). So it would be rather convenient
to develop an efficient and accurate procedure for obtaining the present PDFs
given that for the initial conditions.
Many attempts have been made at solving this problem. Hoffman (1987) 
used the Zel'dovich (1970) approximation for computing the variance
of the density fluctuations in the mildly non-linear regime ($\sigma ^2 \le 1$). However, he
did not take into account the fact that evolution and filtering are non-commuting
operations. To obtain the variance on comoving scale $r$, one should evolve the 
whole field (with all its structure) and then filter it on scale $r$. But in the
analytical procedure followed by Hoffman the field is first filtered and then evolved.
In the linear regime ($\sigma \ll 1$) both procedures lead to the same result. However, for
$\sigma \sim 1$ the results obtained by these procedures are rather different.

A procedure for obtaining the present PDF given the statistics of the initial 
conditions, that
accounts for non-commutability of evolution and filtering was first described
by one of us (Betancort-Rijo 1991). This procedure was the result of taking into account
the simple fact that overdense regions with present comoving size $r$ correspond to
larger comoving regions in the initial field. However, in developing this idea
a couple of subtle points were involved that were not properly dealt within 
that paper.

During more than a  decade that has passed since these first attempts at dealing with the
problem under consideration (the article
 by Betancort-Rijo was in fact written in
1988) much attention have been given to it. New methods have been developed 
for approximating the non-linear evolution (Buchert 1989; Bouchet et al. 1995)
and for the computation of statistics of the evolved field. Bernardeau (1994a)
developed a procedure for taking into account the effect of filtering on the
moments of the PDF in the small variance limit and used it to compute the first 
few hierarchical amplitudes. Then, assuming that these quantities do not depend on $\sigma $, he obtained
(Bernardeau \& Kofman 1995)
the present PDF for any value of $\sigma $. However, this
assumption, although it holds for some initial conditions, is not universal
(as we shall see) and, even when it holds, it is not easy to justify a priori.

Despite  all the attention that has been devoted to the present problem, the
idea described by Betancort-Rijo (1991), which  is at the root of the 
possibility of developing a simple and direct (it is not necessary to obtain the
vertex generation function) procedure for computing  the present PDF in the non-linear 
regime analytically, has not been followed. The exception is an article by Padmanabhan \&
Subramanian (1993), but there the problem is treated in  the same
way as was by Betancort-Rijo (1991), thereby having the same drawbacks.

In this paper we shall show how to correct the errors included in previous
work and give the correct procedure for obtaining the present PDF. To check the
correctness of the procedure we compute the values of $S_3$ and $S_4$ in the 
low-$\sigma $
limit (that have been computed analytically by Bernardeau 1994a) for several power
spectra. Our procedure should render  the exact values of all hierarchical
amplitudes in this
limit, and this is what is found for these two amplitudes. We also study the
dependence on $\sigma $ of these amplitudes and of the variance 
in the low  limit and for finite
values of $\sigma $. We then compute
 the PDF explicitly for several cases for which numerical simulations are available
and find them to agree with the numerical results.

In \S 3 we briefly review the basic idea (fully described in Betancort-Rijo
1991) involved in the present approximation and show how to use it to derive
correctly the present PDF given the initial one. In the second section we shall
discuss the technique that we use for the treatment of the non-linear evolution. In the last section we
present and discuss the results of several explicit calculations.

\section{Non-linear evolution}

  The complete Zel'dovich approximation (CZA; Betancort-Rijo \&
  L\'opez-Corredoira 2000) is an approximation to the non-linear
  evolution of density fluctuations 
  depending only on three quantities, $\lambda _i$, defined so that 
that $1-\lambda _i$ are the eigenvalues of the 
tensor $\frac{\partial \vec{u}}{\partial \vec{q}}$, where
$\vec{u}$ stands for the peculiar velocity of matter as given 
by the linear theory in comoving units, i.e. 
the peculiar velocity $\vec{v}=R(t)\vec{u}$; $R(t)$ being the scale 
factor of the Universe and $q$ stands for the Lagrangian coordinates, which 
are equal to the initial comoving coordinates (Eulerian). This tensor we call 
the linear local deformation tensor. The full local deformation tensor is 
defined in the same manner but using the exact value of $\vec{u}$.
This approximation is exact to the
second order. An approximation exact to the fourth order may be obtained out of it
but this implies introducing additional variables. It takes the form
(Betancort-Rijo \& L\'opez-Corredoira 2000)

\begin{equation}
(1+\delta )^{-1}=\prod _i(1-r_i(\lambda)\lambda _i+
f_1(\lambda )x_i + f_2(\lambda )u_i)
\label{1b}
,\end{equation}
where $f_1$, $f_2$, $r_i$ are certain functions of the
variables $\lambda _i$. The variables $x$, $u$ which are given at any point by certain integrals over the whole
field (Betancort-Rijo 2001) but, for statistical purposes we only need their
 probability distribution for fixed values of the $\lambda _i$.
 In fact, to obtain the
 moments exactly to the third order and almost exactly to the fourth we can use
 expression (\ref{1b}) with $u$ equal to zero and $x$ equal to its average value for fixed  
 values of the $\lambda _i$  $(<x_i>(\lambda ))$. This is the approximation 
 that in principle we use. It has a dependence
 on the power spectra through the mean value of $x$ (see Betancort-Rijo \&
 L\'opez-Corredoira 2000). However, the dependence is so small that may be
 neglected without practical lost of accuracy.  So, what we do in practice is to
 use simply the CZA (expression (1) with $x=u=0$ ), which is strictly exact 
 to the third order for a particular
 power spectra (such that $<x_i>(\lambda )=0$) and practically exact for all of them.
 The CZA gives the density fluctuation at a point in term of the values of 
 the $\lambda _i$ at that point. When dealing with filtered fields, however, the relevant quantity
 is the mean density within certain region (of any prescribed  shape). It may be shown that this mean
 density fluctuation  is also given exactly at least to the third order by
 expression (\ref{1b}) with the $\lambda _i$ representing now the linear
 eigenvalues of the
 deformation tensor corresponding to the filtered field; that is, the
 deformation tensor derived from the peculiar velocity field corresponding to
 the growing mode of the filtered field. The $x$ variables are now defined in a
 different manner. For non-filtered fields the presence of this variables is due to
 the different rate of growth of the various contributions to the local tidal
 field. In the present case, however, these variables have an extra component due
  to the filtering process. But, despite of the different origin of the $x$
  variables, expression (\ref{1b}) and the probability distributions for the $x$  retain
  their form at least up to third order. So,
  setting in this expression $x$ equal to its
  mean value for fixed $\lambda _i$ leads also to an approximation exact to third order
  (for statistical purposes) for filtered fields.
  There is, however, an important difference between the two mentioned
  contributions to the $x$, because the part due to the tidal field remains small
  for any power spectra while that due to the averaging process grows indefinitely
  with increasing power on small scale. This implies, as we shall find, 
  that for
  power spectra with too much power on small scale the procedure presented here
  fails even for values of $\sigma $ smaller than one. However, 
  even in these cases the
  results are exact to third order, but many orders contribute for finite but
  small $\sigma $. So, the steeper the power spectra the smaller the value of
  $\sigma $ at which
  the approximation start failing. This happens, however, only for very steep power
  spectra not very relevant to cosmology. To deal with these cases we should use
  expression (\ref{1b}) with the mean values of $x$, not simply the CZA.

\section{PDF of the filtered field}
    
    The effect of filtering on the field of density fluctuations have been
    treated exactly by Bernardeau (1994). In this treatment one may, in 
    principle, obtaining (previously to caustic formation) all moments of the
    PDF of the field of density fluctuation filtered in some
    scale as a series of powers of $\sigma $ and sometimes (when the corresponding
    Hedgeworth expansion converges) to reconstruct the PDF. However, even when
    this works it is complicated. The procedure described here to deal with this
    problem ,which is basically that in Betancort-Rijo (1991) is more immediate,
    allowing the direct evaluation of the PDF even for values of $\sigma $
    well above unity. Furthermore, it provides a clearer visualization of the effect of
    filtering. We now describe the basic idea of our procedure using the CZA, 
    although the procedure in itself is independent of the approximation used for
    the evolution.
    
    Consider  certain geometrical body, $B_1$, centered at some point within the
    initial field. By assigning to this point the mean value of the density
    fluctuation within $B_1$ and repeating the process for each point, we generate
    a new field which we call the field filtered within $B_1$. We may now obtain
    the linear peculiar velocity field corresponding to the growing mode of this
    field. Let $\lambda _i(B_1)$ be the proper
    values of the corresponding linear (linearly calculated) deformation tensor,
    and let $B_2$ be the body which is the result of transforming $B_1$ with this
    deformation tensor. The CZA, in the present context, must be written in 
    the form:

      \begin{equation}
  1+\delta (B_2)=(\prod _{i=1}^3(1-r_i(\vec{\lambda }(B_1))\lambda _i(B_1)))^{-1}
  \label{4}
  .\end{equation}
   where $\delta $ is the present value of the density fluctuations
    filtered with $B_2$ at the point under
   consideration.  
    In our problem, the filtering body $B_2$ is a sphere of radius $r$ and $B_1$ is an
    ellipsoid with semiaxis $r(1-r_i(\vec{\lambda })\lambda _i)^{-1}$. Hereafter, we
    write simply $\lambda _i$ instead of $\lambda _i(B_1)$. It must be noted that
    in the actual evolution the present sphere does not transform (when evolved
    back) exactly into an ellipsoid. The real surface has the same quadrupolar
    component as the mentioned ellipsoid, but contains smaller wiggles. However,
    for the present problem those wiggles are completely irrelevant.
    
    Equation (\ref{4}) gives the best one-point expression for $\delta $ as
    given by the CZA
    approximation. By one-point expression we mean that the value of $\delta $ at a given
    point (the mean value within a $B_2$ centered at that point) is expressed only in
    terms of the $\lambda _i$  at that point (the mean value within $B_1$). To go beyond this
    approximation, we need to obtain the values of the $\lambda _i$ at different points and
    integrate over them, or, what is the same, to include the x variables
    described in the previous section.
    
    It should be noted that in equation (\ref{4}) the $\lambda _i$ depend on $B_1$, which, in
    turn, depends on the $\lambda _i$ (for a given $B_2$). It is possible to show that, as long
    as caustics do not form, for every point there is a unique ellipsoid (with
    certain axis and orientation) such that the local deformation tensor of the
    field filtered with the ellipsoid transforms that same ellipsoid into a
    sphere of radius $r$ ($B_2$ in general). This prescription uniquely determines  the values of the $\lambda _i$
    (hence, $\delta $) at any given point. These unique $\lambda _i$ (for a given $B_2$) we call $\lambda _i'$. We
    must now obtain the PDF for the $\lambda _i'$ at a randomly chosen point in the initial
    field. With this PDF and equation (\ref{4}), the derivation of the PDF for $\delta $ is a
    standard problem.
    
    For a field filtered with a given body (the same at every point; usually a
    sphere of given radius), the PDF for the $\lambda _i$ is given by (Doroshkevich 1970):
    
    \[
    P(\lambda _1, \lambda _2, \lambda _3, \sigma)=\frac{5^{5/2}27}{8\pi \sigma ^6}(\lambda _1-\lambda _2)
    (\lambda _1-\lambda _3)(\lambda _2-\lambda _3)
    \]\[
    \times \exp\left[\frac{-1}{\sigma ^2}\left(3(\lambda _1+
    \lambda _2+\lambda _3)^2-\frac{15}{2}(\lambda _1\lambda _2+\lambda _2\lambda _3+\lambda _1\lambda _3)\right)
    \right];
  \]\begin{equation}
  \sigma ^2\equiv \langle \delta _L^2\rangle=\langle (\lambda _1+\lambda _2+\lambda _3)^2\rangle;
  \ \lambda _3<\lambda _2<\lambda _1
  \label{5}
  .\end{equation}
 
  $\sigma ^2$ is the variance of the initial field of density fluctuations, $\delta _L$, filtered
      with the given body and extrapolated linearly to the present time. The trace of the deformation tensor extrapolated linearly to the
      present time (is given the tensor used in the Zel'dovich approximation) by  $\delta _L$.
      Betancort-Rijo (1991) and Padmanabhan \& Subramanian
      (1993) took the PDF for the $\lambda _i'$ to be simply proportional to the PDF for
      the $\lambda _i$ (with $\lambda _i=\lambda _i'$) for the field filtered with the ellipsoid that will
      contract to a sphere of radius $r$; that is, with an ellipsoid with semiaxis
      $r(1-r_i(\vec{\lambda '})\lambda _i')^{-1}$.
      Note that now the filtering body changes from one point to
      another, so $\sigma $ is no longer a constant but a function of the $\lambda _i'$.
       
      \[
      \sigma ^2(\vec{\lambda '},r)=\frac{1}{(2\pi )^3}\int 
  	W(F)^2    P(K)d^3\vec{K}; \]
     \begin{equation}
     W(F)=\frac{3[\sin F-F\cos F]}{F^3}
      \label{6}
      \end{equation}
      with $F=\sqrt{\sum _iK_i^2E_i^2}r$; $E_i=(1-r_i(\vec{\lambda '})\lambda _i')^{-1}$, $K=|\vec{K}|$;
      $P(K)\equiv$ power spectra.
  
  In practice, in the papers mentioned and, to some extent, here, instead of the full equation (\ref{6}) a more convenient
      approximation is used. Since $\sigma $ depends mainly on the volume of the filtering
      body (this is exactly so for white-noise power spectra), only the
      dependence on this quantity is considered. We then have:
      
 \begin{equation}
  \sigma ^2(\vec{\lambda '},r)=\sigma ^2(\vec{0},
  r(\prod _{i=1}^3(1-r_i(\vec{\lambda '})\lambda _i')^{-1/3}))\equiv \sigma ^2(
  r(\prod _{i=1}^3(1-r_i(\vec{\lambda '})\lambda _i')^{-1/3}))=\sigma ^2(r(1+\delta )^{1/3})
  \label{7}
  ,\end{equation}
      where $\sigma ^2(r)$ is the variance of the linear field filtered with a sphere of
      radius $r$. 
      However, if we want to obtain results which are exact to the fourth
      order,as we shall do in some cases, the small shape dependence of 
      $\sigma (\vec{\lambda },r)$ must
      be taken into account. In these cases, we use the following excellent
      approximation to the actual shape dependence of $\sigma $:
      
      \[
      \sigma ^2(\vec{\lambda },r)=\sigma ^2(r(1+\delta
      )^{1/3})Z^2(\vec{\lambda})
      \]\begin{equation}
      Z(\vec{\lambda})\approx exp\left[-\frac{B}{2}\sum _{i<j}
      \left(ln\left(\frac{1-\lambda _i}{1-\lambda _j}\right)^2\right)\right]
      \end{equation}

      To compute the constant $B$ in the exponential we may use the exact form of Z
      in the case when two of the $\lambda $ are equal. In this case  and for power law
      power spectra  (with index n) expression (\ref{7}) reduces to:

      \[
      Z(P,n)=\frac{P^{-\frac{2n}{3}}\int _0^\infty \int _0^\infty
      (u^2+P^2s^2)^{\frac{n}{2}}(W(\sqrt{u^2+s^2}))^2 s \ ds\ du}
      {\int_0^\infty W^2(u)u^{2+n}du}
      \]\begin{equation}
      P\equiv \frac{1-\lambda _3}{1-\lambda _1}; \ \ \lambda _1=\lambda _2\ne
      \lambda _3
      \end{equation}

      We find: $B(n=-2)$=0.0486; $B(n=-1)$=0.0499; $B(n=0)$=0.

      The approximation used in the mentioned works (which use expression (5)) may  be expressed
      in the form:
      
\begin{equation}
P'(\vec{\lambda '}, \sigma )=C\ P(\vec{\lambda }, 
\sigma (r(1+\delta (\vec{\lambda }))^{1/3}))|_{\vec{\lambda }=\vec{\lambda '}}
\label{8}
,\end{equation}
where $P'$ is the PDF of the $\lambda _i'$; $\delta (\vec{\lambda })$ is given
by equation (\ref{4}) (with x=u=o); 
$\sigma $ stands for $\sigma (r)$, and $C$ is a normalization constant. This constant has to be introduced
      because, unlike $P$, the integral of $P'$ over all $\vec{\lambda '}$ space is not equal to
      unity, due to the dependence of $\sigma $ on $\vec{\lambda }$.
      
      We  now show why equation (\ref{8}) is incorrect and give the correct one.
      To this end we consider an approximation in which $\delta $ depends only on  
      $\delta _L(\equiv \lambda _1+\lambda _2+\lambda _3)$.      
      We do this because the relevant fact at this point is basically the
       same with one and with three variables, but it is much easier to understand in
       the former case.
        
       For definiteness, let us assume that:
       
\begin{equation}
1+\delta =\left(1-\frac{\delta _L}{3}\right)^{-3}
\label{9}
,\end{equation}
           an approximation that has been used in Betancort-Rijo (1991).
	   
      For Gaussian initial conditions, $\delta _L$, which corresponds to a fixed filtering
      scale, follows a normal distribution. To obtain the PDF for 
      $\delta _L'$ (the
      value of $\delta _L$ within the sphere of radius $r/(1-\frac{\delta
      _L}{3})$ that have contracted at present
      to the sphere of radius $r$) that, by means
      of equation (\ref{9}), allows us to obtain the PDF for 
      $\delta $, we proceed as
      follows. To say that at a given point $\delta _L'$ takes the value 
      $\delta_L^0$ means that at this
      point the linear value of $\delta $ filtered with a sphere of radius $r/\left(1-\delta _L/3\right)$
      is just $\delta _L^0$. The  accumulated probability, $P'(\delta _L'\ge \delta _L^0)$, that at a point chosen at random in
      the initial field $\delta _L'$ is larger than or equal to some given value
         may be obtained immediately. To this end we only need noting that if a
	 sphere of radius $r/(1-\delta _L'/3)$ (with $\delta _L'\ge \delta _L^0$) 
	 have to contract to a
	 sphere of radius r and shell crossing does not happen, the sphere of
	 radius $r/(1-\delta _L^0/3)$ will have to contract to a radius smaller
	 than $r$.
	 So, the value of $\delta _L$ within the sphere with radius
	 $R/(1-\delta_L^0/3)$ must be
	 larger than $\delta _L^0$, since if the value of $\delta _L$ 
	 within this sphere were just $\delta _L^0$
	     it would contract to a sphere of radius r. Thus, the accumulated
	     probability under consideration must be equal to the probability
	     that the value of $\delta _L$ within a sphere of radius 
	     $r/(1-\delta _L^0/3)$ be larger
	     than or equal to $\delta _L^0$. So, we have

      \[
      P'(\delta _L'\ge \delta _L^0)=\frac{1}{2}{\rm erfc}\left(\frac{\delta _L^0}{\sqrt{2}\sigma (\delta _L^0)}\right)
      \] \begin{equation}
      \sigma(\delta _L^0)=\sigma \left(r\left(1-\frac{\delta _L^0}{3}\right)^{-1}\right)
      \label{10}
      \end{equation}
      where erfc stands for the error complementary function. To obtain the
      probability distribution      for the values of $\delta _L'$ (the 
      value of $\delta _L$  within
      the sphere that will contract to a sphere of radius $r$) at a randomly
      chosen point we only need deriving the cumulative probability with respect
      to $\delta _L^0$. 
      
\[
P'(\delta _L')=-\frac{d}{d\delta _L^0}P'(\delta _L>\delta _L^0)|_{\delta _L^0=\delta _L'}
;\]\begin{equation}
P'(\delta _L')d\delta _L'=\frac{1}{\sqrt{2\pi }}e^{-\delta _L'^2/(2\sigma ^2(\delta _L'))}
d\left(\frac{\delta _L'}{\sigma (\delta _L')}\right)
\label{11}
.\end{equation}      
      
      Note that since $\sigma $ depends on $\delta _L'$ we cannot take it out of the differential. So
      we cannot write:
      
\[
P'(\delta _L')=C\ P(\delta _L, \sigma (\delta _L))|_{\delta _L=\delta _L'}
\]\[
=C\frac{e^{-\delta _L'^2/(2\sigma ^2(\delta _L'))}}{\sqrt{2\pi }\sigma (\delta _L')}
,\]    
      contrary to the assumption in equation (\ref{8}). With these considerations in
      mind, it is easy to obtain the correct distribution for the $\lambda _i'$. We need only
      to generalize the above argument to three dimensions. The key argument is
      that the variables $\lambda _i'/\sigma (\vec{\lambda '})$ must follow the distribution followed by the variable
$\lambda _i/\sigma $ for  fixed filtering (when $\sigma $ is a constant). So we have:
	
\[
P'(\vec{\lambda '}, \sigma )=\left(P(\vec{\lambda }, \sigma (\vec{\lambda}))\sigma ^3(\vec{\lambda })
\left| \frac{\partial \frac{\lambda _i}{\sigma (\vec{\lambda })}}{\partial \lambda _j}\right|
\right)|_{\vec{\lambda }=\vec{\lambda '}};
\]\begin{equation}
\sigma (\vec{\lambda })\equiv \sigma (\vec{\lambda },r)
\label{12}
.\end{equation}

	The last factor in this expression is the Jacobian of the transformation
	from the variables $\lambda _i/\sigma (\lambda _i)$ to the variables $\lambda _i$. 
	Note that the $P'$ obtained by
	means of this expression is automatically normalized.
	
	Equation (\ref{12})  is exact for any initial conditions (provided that
	the $\lambda _i$ determine the evolution), although we shall
	use it only for Gaussian initial conditions; that is, with $P$ given by
	equation (\ref{5}).
	
	$P'$ is the PDF of the $\lambda _i'$ (the $\lambda _i$ associated with the unique ellipsoid that
	will contract to the sphere  of radius r centered at the given point) for a point
	chosen at random in the initial field; that is, in Lagrangian
	coordinates. However, we want to obtain the PDF for $\delta $ at a point chosen at
	random in the present ``physical'' coordinates (i.e. Eulerian
	coordinates). So, we must first obtain the PDF for the $\lambda _i'$ in Eulerian
	coordinates. To this end, we only need to multiply $P'$ by the Jacobian of the
	transformation from Lagrangian to Eulerian coordinates. For a
	non-filtered  field
	this Jacobian is simply
	 
\begin{equation}
\prod _{i=1}^3(1-r_i(\vec{\lambda })\lambda _i)
\label{13}
.\end{equation}  
	
	However, for filtered  fields  this expression is not correct. This can
be appreciated by noting that the resulting distribution for the $\lambda _i'$ (the result of
	multiplying $P'$ by the Jacobian in question) is not normalized. For
	non-filtered fields, on the
	other hand, it is easy to check that the Jacobian given by equation (\ref{13})
	leads to a normalized PDF.
	The correct Jacobian must exhibit the following properties: it has to be
	zero at points where $\delta $ goes to infinity; it must reduce to expression
	(\ref{13}) in the limit of small power on scales below $r$; it must render a
	normalized PDF with vanishing first moment for any power spectra. The
	following expression meets these requirements and is exact at least to
	first order

	\begin{equation}
	J(\lambda )=\prod _i(1-(1-\gamma )r_i(\vec{\lambda })\lambda _i)
	\label{14b}
	,\end{equation}
	where $\gamma $ is certain spectral constant that tends to zero in the limit of
	small power on scales below $r$. We cannot say a priori whether this
	expression is exact to higher orders. However,it seems to be exact to
	second order (within the precision of our computations) and it gives
	excellent results to all orders. It must be noted that expression
	(\ref{14b}) is
	not the value of the Jacobian at a point with the given $\lambda _i$. 
	This would be
	so if the evolution were determined solely by these
	quantities. However, we know that this it is not strictly true, so
	expression (\ref{14b}) should represent the mean value of the Jacobian over all
	points with a given value of the $\lambda _i$.
	To obtain $\gamma $ we have considered a fluctuation with spherical symmetry and the evolution
	of the region such that the mean density fluctuation within a sphere
	centered at points belonging to it is above some threshold. Using the
	spherical model with the mean density profile, we obtained the
	contraction of this region and by comparison with expression (\ref{14b})
	we find $\gamma$. However, this derivation is lengthy and although it might be
	  interesting in itself, it is not essential to the present purposes, since we may obtain that constant using any relevant analytical
	  result available (i.e., the leading order value of $S_3$) and comparing it
	  with the one obtained using expression (\ref{14b}). In either case, we find:
	  
	 \begin{equation}
	 \gamma=\frac{1}{6}\frac{d[ln(\sigma ^2(r))]}{d[ln(r)]} 
	  \end{equation}

\section{Results and conclusions}	  
	  
	  To check the accuracy of the approximation developed in this article, we
	  shall present some explicit calculations and compare them with the
	  results of the corresponding available  numerical simulation and with
	  exact results obtained with the perturbation theory. We shall consider Gaussian initial conditions with power
	  spectra of the form $P(k)=Ck^n$.
	  
	  In these cases, equation (\ref{7}) takes the form:

\[\sigma (\vec{\lambda }, r)=\sigma _0(1+\delta )^{-\alpha /2}
;\]\begin{equation}
\sigma _0\equiv \sigma (r); \alpha \equiv \frac{n+3}{3}
\label{30}
,\end{equation}
 	  with $\delta $ given by the CZA

	  \begin{equation}
	  (1+\delta )^{-1}=\prod _i(1-r_i(\lambda )\lambda _i)
	  \end{equation}

	  To derive exact results we use for $r_i(\lambda )$ 
	  its exact expression
	  (Betancort-Rijo \& L\'opez-Corredoira 2000). To derive result for finite 
	   values, however, we find it much more convenient to use the following
	   compact approximation
	   
\[
r_i(\lambda )=g\left(\frac{3}{2}(\lambda _J+\lambda _K),
\Omega\right)	   
\]
\begin{equation}
g(x,\Omega)=\left \{ \begin{array}{ll}
\frac{3}{x}\left(1-\frac{x\Omega ^{2/63}}{1.66}\right)^{0.5533}
& \mbox{;$x\le 1.584$} \\
1.55 & \mbox{;$x> 1.584$}
\end{array} \right \}
\end{equation}

	   Using this in expression (12),
	  with $P(\vec{\lambda }, \sigma )$   
	   given by (3) and multiplying it by the Eulerian--Lagrangian Jacobian, 
	   $J(\vec{\lambda }, \alpha )$, (expression (14)), 
	   we obtain the PDF for the $\lambda _i'$  
	  corresponding to a point chosen at random in Eulerian coordinates, which
	  we represent by $G(\vec{\lambda '}, \sigma _0)$:

\begin{equation}
G(\vec{\lambda '}, \sigma _0)=\left(\left(P(\vec{\lambda },\sigma (\vec{\lambda }))
\sigma ^3(\vec{\lambda })\left|\frac{\partial \frac{\lambda _i}{\sigma (\lambda )}}{\partial \lambda _j}
\right|\right)_{\vec{\lambda }=\vec{\lambda '}}\right) J(\vec{\lambda '}, \alpha)
\label{31}
.\end{equation}

	     The moments of the PDF of the present value of the density
	     fluctuation top-hat filtered on scale r are then given by:
	     
\begin{equation}
\langle \delta ^n\rangle (\sigma _0)=\int _{-\infty}^\infty d\lambda _1'\int_{-\infty}^{\lambda _1'}
d\lambda _2'\int_{-\infty}^{\lambda _2'}d\lambda _3' G(\vec{\lambda '}, \sigma _0)
\delta ^n
\label{33}
,\end{equation}
	     where $\delta $ is given by equations (17) with  
	     $\vec{\lambda }=\vec{\lambda '}$, provided that
	     each of the factors in (17) is positive and $\delta $ is
	     smaller than 170. If
	     this expression leads to a value of $\delta $ larger than this, we use the
	     following expression:
	     
\[
\delta=170\left|\frac{\delta _L}{1.589}\right|^3	     
.\]	     
	     
	     This last prescription represents an attempt to deal with
	     caustic formation in a simple manner, so that equation (23) gives
	     sensible results for any value of $\sigma $. However, it is not relevant for
	     the results given here.
	     
	     The PDF for $\delta $, $P(\delta , \sigma _0)$, may be obtained in the very process of computing
	     any of the moments. To this end, we add up all the values of $G(\vec{\lambda}, \alpha )$
	     (multiplied by the corresponding volume of the integration step)
	     that are found during the integration to lie between $\delta $ and $\delta +\Delta \delta $ (for a
	     suitably chosen $\Delta \delta $, depending on $\delta $) and divide the result by $\Delta \delta $.
	     
	     The values of $\delta $ given by (17) for very under dense regions is not
	     correct this imply $\langle \delta \rangle\ne 0$. 
	     However, this problem disappears if we
	     set $P(\delta )=0$ for $\delta <-0.82$.
	     
	     The values of $S_3$ and $S_4$ given by
	     
\[
S_3\equiv \frac{\langle \delta ^3\rangle}{\langle \delta ^2\rangle ^2};\ \ 
S_4\equiv \frac{\langle \delta ^4\rangle-3\langle \delta ^2\rangle ^2}
{\langle \delta ^2\rangle ^3}
\]	     
	     in the low-$\sigma $ limit, may be computed analytically. 
	The exact value of $S_3(\sigma =0)=(34/7-(n+3))$ 
	     is obtained, for the power spectra under
	     consideration, when expression (17) for $\delta $ is used; 
	     the value $4-(n+3)$ is
	     found when the $r_i$ are set equal to one. For a 
non-filtered field we have computed $S_4(\sigma =0)$ 
analytically both for the CZA (equation (17)) and for the Zel'dovich 
approximation. Note that the non-filtered case may be
 treated formally as a filtered field with negligible power below the filtering
 scale, so that we may use expression (30) with $\sigma $ independent of
 $\lambda $  and $\gamma=0$. For
 power law power spectra this corresponds to the limit $n=-3$, but it corresponds
 more generally to any power spectra having the mentioned properties. The correct values (60712/1323; 279/9; Bernardeau
	     1994a) are found. 
	     For the evolution of these amplitudes and the variance in the small
	     $\sigma $ limit we find
	        
\[
\langle \delta ^2\rangle=\sigma ^2+1.81\sigma ^4+3.8\sigma ^6
\]\[
S_3=4.86+9.815\sigma ^2+48\sigma ^4
\]\begin{equation}
S_4=45.89+288.8\sigma ^2+2072\sigma ^4
\end{equation}

	       This results are exact to next leading order for $S_3$ (in
	       principle only for $n=-2.7$ but the $n$ dependence is negligible) and
	       almost exact to the same order for $\langle \delta ^2\rangle $,
	       $S_4$. As pointed out in
	       section 2, we could use the statistical properties of the $x$,
	       $u$ variables to obtain these two quantities exactly to the stated
	       order, but the second coefficient in the expansion for $\langle
\delta ^2\rangle $  will
	       increase by less than a per cent which is about the numerical error
	       for 
	       this coefficient. So, it would be an unnecessary complication to
	       use here those additional variables. Furthermore, the purpose of
	       the present procedure is to obtain directly and accurately the
	       PDF for finite values of $\sigma $. 
	       The question of the maximum order of
	       exactness, although interesting, is only of peripheral relevance
	       to our purposes.
	       These results agree with those exact results given by the perturbation
	       theory by Scoccimarro \& Frieman (1995) and the remaining ones are close to those  obtained by
	       Fosalba and Gazta\~naga (1998) using the spherical model. We see
	       that this model predict well the hierarchical amplitudes but not
	       so well the moments.
	       The above results provide a check to the CZA but not properly to
	       the procedure presented here, since for non-filtered fields
	       expressions (12) and (14) reduce to the standard form. A check to
	       these expressions is provided by the cases of power law power
	       spectra with $n\ge-3$. For $n=-2$ we find:

\[\langle \delta ^2\rangle=\sigma ^2+0.88\sigma ^4+0.54\sigma ^6
\]\[
S_3=3.86+3.12\sigma ^2
\]\begin{equation}
S_4=27.56+58\sigma ^2
\end{equation}

	       Here, to obtain the second coefficient in $\langle \delta
	       ^2\rangle $ exactly we have
	       considered the small (=0.015) contribution due to the
	       dispersion of the $x$ variables. This is not necessary for the
	       amplitudes. It is also important to consider the shape dependence
	       of $\sigma $ (see expression (6)), otherwise we would obtain a coefficient 0.1 too
	       large. For $n=-1$, $n=1$ we find respectively:

\[\langle \delta ^2\rangle=\sigma ^2+0.223\sigma ^4
\]\[
S_3=2.86+0.267\sigma ^2
\]\[
S_4=13.89-2.583\sigma ^2
\]

\[\langle \delta ^2\rangle=\sigma ^2-0.173\sigma ^4
\]\[
S_3=0.86+0.8\sigma ^2
\]\begin{equation}
S_4=0.556-17.8\sigma ^2
\end{equation}

	     For $n\ge -1.5$, $S_3$ and $S_4$ are practically independent of $\sigma $. For example,
	     for $n=-1.3$ the change of these quantities up to $\sigma =1.52$ is less than a 4\%. For
	     smaller values of $n$, however, the evolution of this quantities is more
	     apparent. For $n=-2$ we find (table 1) that $S_3$ and $S_4$ increase
	     considerably from $\sigma =0$ to $\sigma \approx 0.5$ and more steeply for higher values of $\sigma $.
	     In figure 1 we present the PDF for $\delta $ corresponding to $P(k)\propto k^{-1}$ and
	     $\sigma =0.3;\ 0.5$. Comparing this with figure 7  of
	     Bernardeau \& Koffman (1995), we find them to be almost identical.
	     The PDFs of Bernadeau \& Koffman were  obtained under the
	     assumption that the $S_p$ parameters (the $p$-th cumulants divided by $\sigma ^{2(p-1)}$)
	     are independent of $\sigma $. We have seen that this is the case for $n=-1$, at
	     least for $S_3$ and $S_4$. So, the agreement between both pairs of PDFs is
	     to be expected.
The evolution of $\langle \delta ^2\rangle$ agrees very well with the result
by Lokas et al. (1996).
	     
	     In figure 2 the PDFs corresponding to $\sigma =0.92;\ 1.52$, $n=-1.3$ are
	     represented. These should be compared with figure 10 of Bernardeau
	     \& Kofman (1995). The agreement is very good, our results being
	     always within the numerical errors of the results given in that
	     article.
	     
	     It should be noted that the mentioned numerical results assume a cold
	     dark matter (CDM) while we have used a power-law power spectrum
 with the same local logarithmic
	     derivative. Using the whole CDM power spectrum in our computations
	     would represent only a small complication. If we have not done so, it
	     is because this implies only a slight improvement in the
	     results that is completely unobservable given the precision of the
	     simulations.
	     
	     The procedure described here for obtaining the PDF for $\delta $ is in
	     principle valid only for top-hat filtered fields. However, the basic
	     lines of reasoning followed here are entirely translatable to 
	     another smoothing procedures.
	     
	     For a given smoothing (in the present field), we must determine the
	     smoothing in the initial field rendering  the minimum dispersion for
	     the smoothed present field. That is, we must find a smoothing
	     procedure  that, when applied to the initial field at a given point,
	     renders a set of $\lambda _i'$ (extrapolated linearly to the present time) that
	     predict the mean value of $\delta $ at the transformed point, given by an
	     expression similar to (17), and so that the dispersion in $\delta $ for given $\lambda _i'$s
	     is a minimum. This is what we have implicitly done here for the
	     spherical top-hat smoothing, and we have found that it corresponds to an
	     initial ellipsoidal top-hat smoothing.
	     
	     The solution to this problem for  Gaussian smoothing is not simple,
	     although one could guess that a triaxial Gaussian, with the axes
	     determined as in the top-hat case, should not be a bad
	     approximation. Within this simple approximation and for a power-law
	     spectrum the PDF for $\delta $ must be the same as for the top-hat case with
	     the same $\sigma $. For more general power spectra, we cannot decide a
	     priori whether the value of the logarithmic derivative must be taken
	     on the scale where the top-hat has the same volume as the Gaussian
	     or on the scale where it has the same $\sigma $.
	     
	     As an example we discuss one of the cases represented in figure 4 
of Kofman et al.  (1994), where the Gaussian smoothing radius
	     is $R_s =6 h^{-1}$Mpc; $\sigma =0.55$. If we matched the volumes, the corresponding
	     top-hat radius is 9.38 $h^{-1}$Mpc, so that for a CDM power spectrum
 $n\approx -1$. The
	     PDF corresponding to a top-hat filtering with $\sigma =0.55$; $n=-1$ is      
	  represented in figure 3 and shows very good agreement with the
	     numerical results in the last reference. If we matched the $\sigma $'s, the
	     corresponding top-hat smoothing scale imply $n=-1.1$. The corresponding
	     PDF is almost indistinguishable from the former.
	     
	     We have seen that to obtain the PDF for $\delta $ we must integrate over all values 
	     of $\lambda $
	      subject to the condition $\delta (\lambda )=\delta $. If we considered, however, an
	     approximation where $\delta $ depended on the $\lambda _i$ only through $\delta _L=
	     \lambda _1+\lambda _2+\lambda _3$, the
	     computation of $P(\delta )$ would be greatly simplified. We have already seen
	     that such an approximation cannot be entirely correct, but  leads
	     to a simple analytical expression that displays, in a clear manner,
	     the effect of smoothing. Taking for $\delta (\delta _L)$ that corresponding to 
	     spherical collapse and using the fact that the distribution for $\delta _L$ is
	     simply a Gaussian, equation (\ref{11}) and (14) immediately gives  for $P(\delta )$:
	     
\[
P(\delta )=\frac{1}{\sqrt{2\pi }}\frac{e^{-\frac{\delta _L^2(\delta )}{2\sigma ^2(\delta )}}}
{\sigma (\delta )(1+\delta )}\left((1+\delta )^{-0.602}+\frac{\alpha }{2}
\delta _L(\delta )\right)
\]\[ \times
\left(1-(1-\frac{\alpha	}{2})\frac{\delta _L(\delta )}{1.66}\right)^{1.66};
\]\[
\sigma (\delta )=\sigma _0 (1+\delta )^{-\alpha /2}; \ \ \ \alpha \equiv \frac{n+3}{3}
\]\begin{equation}
\delta _L(\delta )=1.66(1-(1+\delta )^{-0.6024})
\label{27p}
.\end{equation} 
 
 Where the last equation represents a very accurate approximation to the
 spherical model that we propose.
	     However, expression (\ref{27p}) represents only a rough
	     approximation to $P(\delta )$. For $\sigma =1.52$ the value of
	     $P(\delta )$ given by (\ref{27p}) may differ from the real value by a factor greater
	     than three. Furthermore, the value of $S_3(\sigma =0)$ given by (\ref{27p})
	     is $37/7-(n+3)$, which is 3/7 too large.
	     
	     We may conclude that for $\delta \in (-0.9,10)$ and $\sigma $ as large as 1.52, the
	     $P(\delta )$ computed as indicated here is, so far as we have been able to
	     check, in excellent agreement with the results of numerical
	     simulations. This agreement is particularly meaningful since our
	     derivation of $P(\delta )$ involves neither free parameters nor modeling
	     freedom (except for some minute refinements non-essential for the
	     results given here). Equation (\ref{12}) is exact and is valid for any
	     field. Equation (14) for the Jacobian of the Eulerian--Lagrangian
	     transformation in a filtered field is also exact (at least to
	     first order in $\lambda _i$) and, although derived for a Gaussian
	     field, it seems to be valid for any field.


\eject

{\bf FIGURE CAPTIONS}

Fig. 1: Probability distribution function (PDF) for $\sigma =0.3$ and 
$\sigma =0.5$; $n=-1$.

Fig. 2: Probability distribution function (PDF) for
$\sigma =0.92$ and $\sigma =1.52$; $n=-1.3$.

Fig. 3: Probability distribution function (PDF) for 
$\sigma =0.55$; $n=-1$.

\begin{figure}[htb]
\label{Fig:1}
\end{figure}
\begin{figure}[htb]
\label{Fig:2}
\end{figure}
\begin{figure}[htb]
\label{Fig:3}
\end{figure}
%
%
%
%
%
%

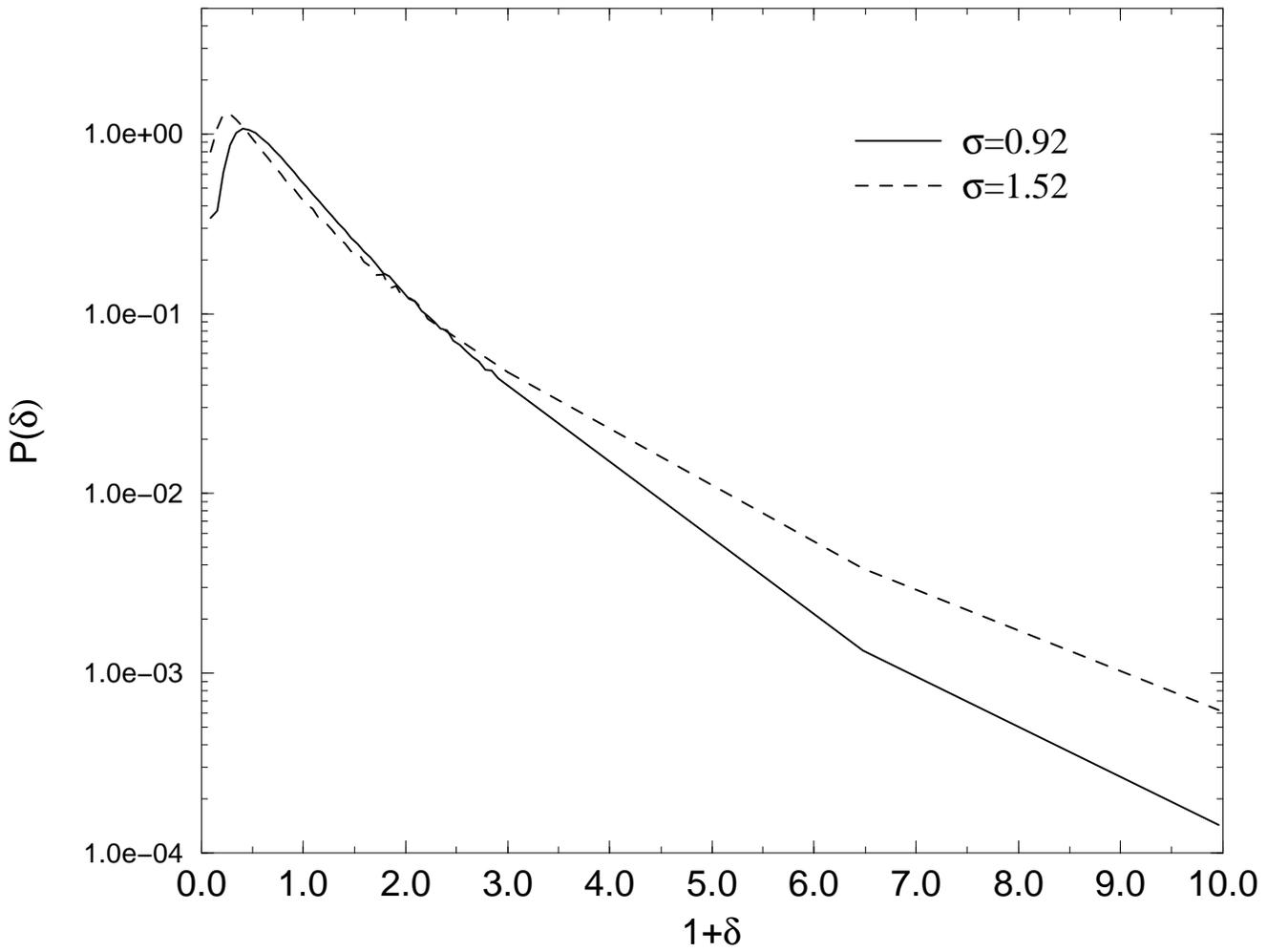
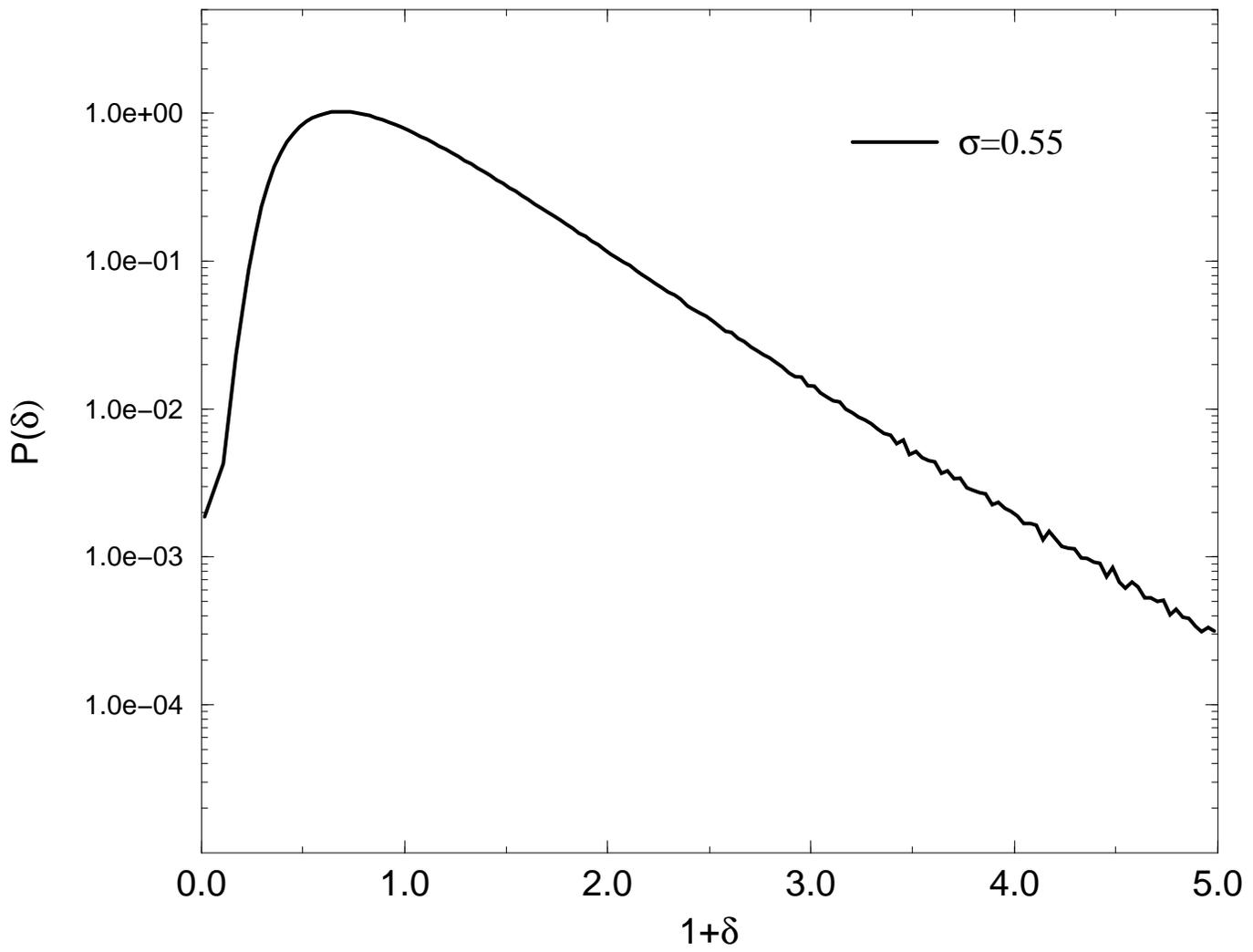
\end{document}